# User-Defined Operators Including Name Binding for New Language Constructs

Kazuhiro Ichikawa[a] and Shigeru Chiba[a]

a    The University of Tokyo, Japan

**Abstract**    User-defined syntax extensions are useful to implement an embedded domain specific language (EDSL) with good code readability. They allow EDSL authors to define domain-natural notation, which is often different from the host language syntax. Recently, there have been several research works of powerful user-defined syntax extensions.

One promising approach uses user-defined operators. A user-defined operator is a function with user-defined syntax. It can be regarded as a syntax extension implemented without macros. An advantage of user-defined operators is that an operator can be statically typed. The compiler can find type errors in the definition of an operator before the operator is used. In addition, the compiler can resolve syntactic ambiguities by using static types. However, language constructs involving static name binding are difficult to implement with user-defined operators. Name binding is an association between names and values (or memory locations). Our inquiry is whether we can design a system for user-defined operators involving a new custom name binding.

This paper proposes a variant of a lambda expression, called a *context-sensitive expression*. A context-sensitive expression looks like a normal expression but it implicitly takes parameters. To realize user-defined name binding, these parameters are accessed through public members such as methods and operators of a parameter object instead of through parameter names. Programmers can emulate name binding by such methods or operators, for example, they can emulate a local variable through a getter operator and a setter operator. We also present *turnstile type* to give a type to a context-sensitive expression, *DSL class* to cooperate with syntax extension, and *generic name* to recognize arbitrary names given by users.

A contribution of this paper is that we reveal that we can integrate a system for managing names and their scopes with the module system and type system of an object-oriented language such as Java. Programmers can define new language constructs involving custom name binding and other programmers can use them by importing modules. The compiler can statically verify the types of both the definition and usage of the language constructs.

We have implemented *ProteaJ2*, a programming language based on Java, and it supports our proposal. We describe a parsing method that adopts *eager disambiguation* for fast parsing and discuss its time complexity. To show the practicality of our proposal, we have conducted two micro benchmarks to see the performance of our compiler. We also show several use cases of DSL classes for demonstrating DSL classes can express various language constructs.

Our ultimate goal is to let programmers add any kind of new language construct to a host language. To do this, programmers should be able to define new syntax, name binding, and type systems within the host language. This paper shows that programmers can define name binding with their own syntax.

# The Art, Science, and Engineering of Programming

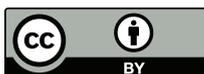

**User-Defined Operators Including Name Binding for New Language Constructs**

# 1 Introduction

Currently, programmers can implement libraries that look like new language constructs. These libraries can be used as if they were built-in language constructs [26]. Modern languages such as Scala and Ruby provide features that enable us to define such libraries. Programmers can implement these libraries by defining functions, methods, or operators so that they will appear to be a built-in construct. For example, we can define unless in Scala as follows:

```
def unless (cond: Boolean)(body: => Unit): Unit = { if (!cond) body }
```

We can use it as follows:

```
unless (a < 100) { println("a is smaller than 100") }
```

Our unless looks like a built-in language construct if. User-defined language constructs can make a program shorter and hopefully easier to understand. In particular, they are useful for improving the readability of code written in an embedded domain specific language (EDSL).

Syntactic macros [9, 14] are well-known functionality for implementing new language constructs. A syntactic macro is a translation rule from one abstract syntax tree (AST) to another. We can implement a new language construct by defining a syntactic macro that translates the language construct into a code snippet that uses predefined language constructs. Syntactic macros are widely used in Lisp since a Lisp program consists of only S-expressions. Programmers can easily modify an AST since the AST is very simple. Furthermore, a macro call looks like a built-in language construct. Syntactic macros have also been implemented in non-Lisp languages, such as Java and Haskell [1, 21].

Although syntactic macros themselves do not extend syntax, they are sometimes used as the back-end of a syntax extension system [6]. Nemerle [23] provides a powerful syntax extension system based on syntactic macros. For example, we can define a language construct foldfor as follows:

```
macro FoldForMacros (id1, r, id2, list, body)
syntax ("foldfor", "(", id1, "assign", r, ";", id2, "of", list ")", body) { <[
  mutable $id1 = $r;
  foreach ($id2 in $list) { $body }
  $id1
]> }
```

Line 2 in this definition declares the syntax of the language construct. The first token foldfor is a unique identifier to distinguish this syntax from other user-defined syntax. The rest of the syntax consists of operators or identifiers expressed by string literals and the names of parameters. The semantics of the syntax is defined by a syntactic macro. Lines 3 to 5 define the macro expansion rule. This syntax can be used as follows:

```
def sumOfSquare = foldfor ( a assign 0 ; i of xs ) { a = a + i * i }
```

Syntactic macros are not the only approach to syntax extension systems. Another approach is to use user-defined operators. A user-defined operator is a function that has its own syntax. In Scala, for example, programmers can define an operator that has arbitrary binary infix syntax. An advantage of user-defined operators over user-defined syntactic macros is that an operator can be statically typed. The compiler can find bugs in the definition of an operator before the operator is used. Our previous work, called ProteaJ [12], is an extended Java language. It provides user-defined operators that can express the arbitrary syntax of parsing expression grammars (PEGs) [8]. Although ProteaJ can express various kinds of syntax found in many programming languages, it cannot express language constructs such as foldfor. This is because ProteaJ user-defined operators cannot support name binding. Name binding is the association of names with values. A user-defined operator is just a function with custom syntax and hence it does not have the ability to customizing name binding. In fact, a compiler resolves name bindings after the parsing or syntactic analysis processes are complete. While the resolution of name bindings is not part of parsing, a user-defined operator in ProteaJ is just a mechanism for customizing the parser.

To enable a user-defined operator with custom name binding, this paper proposes *context-sensitive expressions*, which are a variant of lambda expressions. A context-sensitive expression looks like a normal expression but it is a lambda-like expression implicitly taking parameters. These parameters are accessed through public members such as methods or operators of one parameter object instead of through parameter names. We can emulate name binding by using such public members; for example, we can emulate a variable named "it" by a nullary operator "it" and a unary operator "it = _". We have developed ProteaJ2, which is a programming language based on Java but supports context-sensitive expressions. ProteaJ2 supports *turnstile types*, *DSL classes*, and *generic names* to express context-sensitive expressions for syntax extension with name binding. To efficiently parse a program using a context-sensitive expression, the parser of ProteaJ2 adopts *eager disambiguation* using expected types.

In the rest of this paper, we first show existing approaches for implementing user-defined language constructs with custom name binding. Then we propose *context-sensitive expressions* and present our language design for supporting them in Section 3. Section 4 explains a parsing method and the restrictions of our language for efficient parsing. In Section 5, we discuss the time complexity of our parsing method and show several examples to demonstrate the expressiveness of our proposal. Section 6 is on related work. We conclude in Section 7.

## 2 Motivation

### 2.1 User-Defined Operators

Powerful user-defined operators can be used for implementing user-defined language constructs with a user-friendly syntax. For example, the following code uses a user-defined language construct "p":

```
1   p "hello, world!";
```

**User-Defined Operators Including Name Binding for New Language Constructs**

This code displays "hello, world!" to the user. The language construct "p" is defined as a unary operator as follows:

```
void "p" _ (String s) { System.out.println(s); }
```

This is a code snippet in ProteaJ [12], which is a language proposed in our previous work. ProteaJ has Java-like syntax and supports powerful user-defined operators. In this paper, as in ProteaJ, we assume that we can define an operator with an arbitrary name like "p". The above code snippet is a definition of a user-defined operator. void at the beginning of the snippet is the return type of the operator. The following "p" _ indicates the syntax of the operator. The double-quoted string "p" expresses a *name part* of the operator and the underscore _ expresses an *operand* of the operator. A name part is a symbol for identifying the operator. An operand takes an expression with the corresponding type that is specified in the parameter list of the operator. The parameter list is a parenthesized part following the syntax definition. According to the parameter list, the corresponding type of the operand is String. The curly braced code is an operator body that is evaluated when the operator is called.

An advantage of user-defined operators over user-defined syntax extensions based on macros is that an operator can be statically typed. The compiler can find type errors in the definition of an operator before the operator is used. In addition, it allows for another useful feature: the compiler can resolve syntactic ambiguities by using type information. User-defined syntax extensions such as user-defined operators often conflict with another syntax extension. This is because the authors of different syntax extensions do not know about extensions created by each other. In such a case, the parser might generate multiple valid abstract syntax trees (ASTs) for single fragment of source code. We call such source code ambiguous. To resolve these ambiguities, the compiler should infer the user's intention: which AST represents the user's intent. The compiler can exploit type information to infer the user's intention.

**2.2 Name Binding**

A disadvantage of user-defined operators is that it is difficult to implement language constructs involving *name binding*. Name binding is the association of data with identifiers. Language constructs, especially control-flow statements, often have name binding. For example, Figure 1 shows a code snippet using user-defined operators, which implement control flow statements fold-for and if-exists. This code snippet traverses a list and calculates the number of occurrences of each identical item in the list. The control flow statement fold-for introduces two names, acc and n, and they are available in the body of fold-for. acc expresses an accumulator and n expresses an element of the given list. Here, acc is a Map and n is a key of the Map. The initial value of acc is {}, which is an empty Map. The accumulator acc is updated for each element of the given list according to the body of fold-for. The control flow statement if-exists also introduces a name, it. it expresses a reference to the value of acc[n] if it exists. it is available only in the body of if-exists, in which it behaves as a mutable variable that corresponds to the given key n in the given Map acc. Note that if-exists takes three arguments rather than two: acc and n are different arguments of if-exists.

**Figure 1** Code snippet using user-defined control flow statements fold-for and if-exists

```
1  fold—for (acc = {}; n : list) {
2    if—exists (acc[n]) it = it + 1
3    else acc[n] = 1
4  }
```

**Figure 2** Code snippet using fold-for and if-exists represented by higher-order abstract syntax

```
1  fold—for (Collections.emptyMap(), list, (acc, n) —> {
2    if—exists (acc.get(), n.get(), (it) —> {
3      it.set(it.get() + 1)
4    }, () —> {
5      acc.get().put(n.get(), 1)
6    })
7  })
```

It is because if-exists can modify the entry of the given `Map`, `acc`, by the assignment operator it = _.

If we do not need custom syntax, name binding can be expressed using a lambda expression. *Higher-order abstract syntax* (HOAS) [20] is a generalized technique of this idea. For example, the following code snippet using the language construct let:

```
1  let x = 10
2  x = x + 1
3  println(x)
```

is represented in HOAS as follows:

```
1  let (10, (x) —> {
2    x.set(x.get() + 1)
3    println(x.get())
4  })
```

In the HOAS representation, a variable name is encoded into a parameter name of a lambda expression and the variable value is given by the application of the lambda expression. If we use HOAS representation, we can represent arbitrary language constructs having name binding. For example, Figure 2 shows the HOAS representation of a code snippet in Figure 1. According to line 1 of Figure 2, the language construct fold-for is expressed as a function with three arguments. The first argument is the initial value of the accumulator and the second argument is the target list. The last argument is the body of fold-for, which is represented as a lambda function with two arguments. The arguments of the lambda expression expresses bound names, `acc` and `n`. if-exists is also represented in the same manner.

Recaf [2] supports a limited form of user-defined syntax based on HOAS. In Recaf, programmers can define a new language construct with one of the supported forms of syntax such as syntax similar to `for` or `while` statement. The new language construct is transformed into its HOAS representation. The programmer only implements a method that corresponds to the name of that language construct. The following code snippet is a definition of the let statement in Recaf, which has the same syntax as for statement:

```
public <U> IExec Let (ISupply<U> v, Function<U, IExec> body) {
    return (s) —> { body.apply(v.get()).exec(null); };
}
```

The let statement can be used as follows:

```
let (String s : "hello, world!") {
    System.out.println(s);
}
```

Unfortunately, Recaf cannot translate every possible syntax into a HOAS representation. To transform a code snippet into a HOAS representation, the compiler has to know which code snippets introduce names and where the names are available. In Recaf, such knowledge only comes from Java, so programmers can only define language constructs that have similar syntax and the same name binding rule as predefined language constructs in Java. The language construct fold-for cannot be defined in Recaf since there is no language construct in Java that has similar syntax and name binding to fold-for.

If we use a syntax extension system based on syntactic macros, we can translate arbitrary syntax into its HOAS representation. However, this approach is known to be dangerous. It is difficult to verify that a syntactic macro always generates correct code. When a syntactic macro causes an error after expansion, programmers cannot know whether the usage of the macro is wrong or the macro definition is wrong. Furthermore, syntactic macros make it difficult to cooperate with external tools such as integrated development environments (IDEs).

## 3 Proposal : Context-Sensitive Expressions

HOAS can represent name binding but its syntax extensibility is weak without support for syntactic macros since HOAS uses lambda expressions as a primitive for representing name binding. We present a variant of a lambda expression, which we call a *context-sensitive expression*, as another primitive for representing name binding. A context-sensitive expression takes several parameters as a lambda expression does but the individual parameters are not explicitly given. They are given through the visible members, such as methods and operators, of a few parameter objects. They are available within the expression as if they are local variables. This enables more natural syntax for name binding. We have developed ProteaJ2, which is a programming language based on Java but supports context-sensitive expressions. ProteaJ2 supports *turnstile types*, *DSL classes*, and *generic names* to express context-sensitive expressions for syntax extension with name binding. The source code of the compiler is available from our github repository.[1]

---

[1] https://github.com/csg-tokyo/proteaj2

### 3.1 Context-Sensitive Expressions

A *context-sensitive expression* looks like a normal expression but it implicitly takes parameters. These parameters are accessed through public members of one parameter object. For example, consider the following code snippet:

```
if—exists (acc[n]) it = it + 1
else acc[n] = 1
```

This code snippet is a part of Figure 1. The expression it = it + 1 is an argument of the language construct if-exists. This expression can be regarded as a context-sensitive expression that implicitly takes a reference to each entry of the given Map. This expression can be also written with a lambda expression as follows:

```
(it) —> it.set(it.get() + 1)
```

With a context-sensitive expression, we can write it as follows:

```
set(get() + 1)
```

As you can see this code, the declaration of the parameter "it" and the receiver of set and get are omitted. This is similar to the omission of this (or self) in an instance method. What is different from the this omission is that the implicit receiver is given by the context; here, it is given by if-exists. The code snippet set(get() + 1) does not look natural but programmers can change the syntax of the getter and setter by using user-defined operators. If programmers define the getter syntax is "it" and the setter syntax is "it = _", the code snippet is written as follows[2]:

```
it = it + 1
```

Since a context-sensitive expression is a variant of a lambda expression, the value of a context-sensitive expression is an anonymous function. The language construct if-exists can be defined as following pseudo code:

```
def "if—exists" "(" _ "[" _ "]" ")" _ "else" _ (map, key, thn, els) {
  if (map.contains(key)) thn.apply(new MapEntryRef(map, key));
  else els.apply(new Lazy());
}
```

This implements a new operator. Here, thn is a parameter taking a context-sensitive expression. The value of thn is an anonymous function that takes an argument expressing an implicit receiver of the getter "it" and the setter "it = _", here, a `MapEntryRef` object. thn.apply(...) denotes a call of the anonymous function thn.

Note that a context-sensitive expression is lazily evaluated. If an object passed to a context-sensitive expression does not have a method or an operator, the context-sensitive expression is just for lazy evaluation. The parameter els above is an example of such a context-sensitive expression. A Lazy object passed to els by apply does not have any members, and hence any operators such as "it" and "it = _" are not available

---

[2] "it" can be regarded as a nullary operator and "it = _" as a unary operator taking an assigned value

within the expression that els refers to. The expression is just a normal expression although it is treated as a lambda expression and lazily evaluated.

If context-sensitive expressions are nested, public members of the implicit receiver of the outer expression can also be used in the inner expression. For example, if the language construct fold-for and if-exists are defined by using context-sensitive expressions, we can write the program shown in Figure 1. The body of fold-for is a context-sensitive expression in which the operators acc and n are available and the body of if-exists is also a context-sensitive expression as mentioned above. The operators acc and n are also available in the body of if-exists. If we nest if-exists, multiple identical operators are visible from the inner expression. In this case, the operator of the innermost expression is available there.

### 3.2 Turnstile Types

To give a type to context-sensitive expressions and distinguish from normal expressions, ProteaJ2 provides *turnstile types*. A turnstile type is written as $S \vdash T$, which read as *"a type T under the assumption S"*. A turnstile type cannot be used everywhere; it is only available as a parameter type. If a turnstile type is specified as a parameter type, an argument of the parameter is a context-sensitive expression. For example, the following code defines the language construct if-exists by using turnstile types:

```
<K, V> void "if—exists" "(" _ "[" _ "]" ")" _ "else" _
  (Map<K, V> map, K key, MapEntryRef<K, V> |— Void thn, Lazy |— Void els) { ... }
```

The body is the same as one shown in the previous pseudo code. In this code, the third parameter and the fourth parameter have turnstile types. Assuming `MapEntryRef` defines getter and setter operators, "it" and "it = _", we can write the following program using if-exists:

```
if—exist (acc[n]) it = it + 1
else acc[n] = 1
```

The argument passed to thn is the expression it = it + 1. Because of its type, Void is the type of the resulting value of the expression under the assumption that `MapEntryRef<K,V>` is the type of the implicit receiver object of "it" and "it = _". The compiler will type-check the expression (and also thn.apply(...) in the body of if-exists) to ensure that it satisfies this assumption.

As mentioned above, a context-sensitive expression is an anonymous function. In ProteaJ2, a value of turnstile type $S \vdash T$ is implemented by a function object `Function<S, T>`. For example, thn is an object of type `Function<MapEntryRef<K, V>, Void>` in the body of if-exists. Hence, the context-sensitive expression it = it + 1 is compiled into the following lambda expression:

```
(ref) —> ref.set(ref.get() + 1)
```

In other words, a program with context-sensitive expressions can be regarded as being translated into its HOAS representation using lambda expressions.

■ **Figure 3** The definition of the DSL class MapEntryRef

```
1  dsl MapEntryRef <K, V> {
2    V "it" () { return map.get(key); }
3    void "it" "=" _ (V value) { map.put(key, value); }
4    MapEntryRef (Map<K, V> map, K key) { this.map = map; this.key = key; }
5    private Map<K, V> map;
6    private K key;
7  }
```

### 3.3 DSL Classes

A *DSL class* is a module system for user-defined operators. It is similar to a normal class in an object-oriented language like Java but it contains operators rather than methods. When a DSL class type appears on the left-hand side of a turnstile type, the operators declared in that DSL class are available in the context sensitive expression of that turnstile type. The instance operators are invoked on an instance of the DSL class passed to that expression. Figure 3 shows the definition of the DSL class MapEntryRef. Line 2 defines the getter operator "it" and line 3 defines the setter operator "it = _". These operators are available in an expression of the turnstile type MapEntryRef<K, V> ⊢ R. Line 4 defines a constructor of MapEntryRef. Line 5 and line 6 defines instance fields of MapEntryRef. As a normal Java class, a DSL class is instantiated by new and an instance operator can access instance fields.

A DSL class can also contain static fields and operators. In ProteaJ2, a static operator of a DSL class is available anywhere, not only within a context-sensitive expression. For example, if-exists is a static operator. Figure 4 shows the definition of the DSL class MapUtils, which contains the operator if-exists. It also contains several other operators for accessing a Map object such as _ [ _ ] = _. These operators are available anywhere in a source file that imports the DSL class MapEntryRef. A DSL class is imported by an import dsl declaration. The following code is an example using the operators defined in Figure 4:

```
1  import dsl MapUtils;
2  …
3  Map<String, Color> colors = {};
4  colors["red"] = Color.RED;
5  Color c = colors["red"]; // Color.RED
```

As in Python, we can use the syntax of associative arrays for Map by importing MapUtils. In ProteaJ2, the import dsl declaration is written at the top of the source file (at the same position as an ordinary import declaration).

### 3.4 Generic Names

To recognize arbitrary names given by users, ProteaJ2 provides *generic names*, which extend generics. A generic name is a type parameter of a method or an operator. It is used for checking whether a name is identical to a name that appeared before. For

## User-Defined Operators Including Name Binding for New Language Constructs

■ **Figure 4** The definition of the DSL class MapUtils that contains the operator if-exists

```
1   dsl MapUtils {
2     static <K, V> void "if—exists" "(" _ "[" _ "]" ")" _ "else" _
3       (Map<K, V> map, K key, MapEntryRef<K, V> |— Void thn, Lazy |— Void els) {
4         if (map.contains(key)) thn.apply(new MapEntryRef<K, V>(map, key));
5         else els.apply(new Lazy());
6       }
7     static <K, V> Map<K, V> "{}" () { return new HashMap<K, V>(); }
8     static <K, V> V _ "[" _ "]" (Map<K, V> map, K key) { return map.get(key); }
9     static <K, V> void _ "[" _ "]" "=" _ (Map<K, V> map, K key, V value) { map.put(key, value); }
10  }
```

example, in Figure 1, the language construct fold-for binds two names given by the user, acc and n. These names also appear in the body of fold-for. Generic names are used for recognizing their identity.

Figure 5 shows the definition of fold-for, which uses generic names for dealing with names given by users. The generic names are id1 and id2 in the type parameters of the DSL class FoldFor (line 1) and the static operator fold-for (line 2). Id is the type name of the generic names. The generic names id1 and id2 can appear in an operand part of operator syntax. They specify that all the occurrences of id1 or id2 refer to an identical expression (i.e. name). For example, line 2 and 3 in Figure 5 specifies that the operand id1 of fold-for is identical to the operands id1 of the following operators if they occur in the context-sensitive expression f.

```
1   R id1 () { return acc; }
2   void id1 "=" _ (R r) { this.acc = r; }
```

This is specified by the assumption part of f's type, FoldFor<T,R,id1,id2>. The expression f can invoke the operators declared in FoldFor<...> where id1 is identical to the operand id1 of fold-for. Thus, the invocation of fold-for in Figure 1 is valid. Here, id1 is bound to acc and id2 is bound to n. FoldFor<...> declares the getter operators acc and n and the setter operator acc = _. They are available in the expressions at line 2 and 3 in Figure 1.

The generic names id1 and id2 refer to an expression of type Id. The compiler recognizes that two generic names are identical if the abstract syntax trees of their expressions are equivalent. We chose this design since we express a name by the composition of several user-defined operators. Note that the type Id is not a primitive or meta type. It is a type that users can define for expressing an arbitrary sequence of alphabet letters as an expression. See the following operators for Id values:

```
1   literal Id _ _ (Letter letter, Id rest) { return new Id(letter.toString() + rest.toString()); }
2   literal Id _ (Letter letter) { return new Id(letter.toString()); }
3   literal Letter "a" () { return Letter.a; }
4   …
5   literal Letter "z" () { return Letter.z; }
```

literal is a modifier indicating the operator is for user-defined literals in ProteaJ2. The syntax of a literal operator does not recognize whitespace characters as a separator between name parts and operands. An operand of a literal operator is an expression that consists of only literal operators. User-defined literals defined by literal operators

**Figure 5** The definition of fold-for in ProteaJ2

```
1   dsl FoldFor <T, R, id1: Id, id2: Id> {
2     static <T, R, id1: Id, id2: Id> R "fold—for" "(" id1 "=" _ ";" id2 ":" _ ")" _
3       (R ini, List<T> list, FoldFor<T, R, id1, id2> |— Void f) {
4       FoldFor<T, R, id1, id2> env = new FoldFor<T, R, id1, id2>(ini);
5       for (T t : list) {
6         env.elem = t;
7         f.apply(env);
8       }
9       return env.acc;
10    }
11    R id1 () { return acc; }
12    T id2 () { return elem; }
13    void id1 "=" _ (R r) { this.acc = r; }
14    FoldFor (R ini) { this.acc = ini; }
15    private R acc;
16    private T elem;
17  }
```

can be used as a normal expression. Therefore, if we import the operators above, the following code is valid:

```
1   Id id = acc;
```

The right-hand side of = is the composition of operators. It is parsed like toId(a(), toId(c(), toId(c())))). As in our previous work ProteaJ [12], a and c are recognized as operators since the expected type of the expression is Id. If a generic name refers to an expression like acc, the expression is not evaluated at runtime or at compile time. It is compiled out to be a unique identifier.

### 3.5 Operator Priorities

ProteaJ2 supports *operator priorities* for helping users to define and use operators. Figure 6 shows a definition of the DSL class MapUtils with operator priorities. Line 2 in the figure is a declaration of priorities. This declaration declares three priority names p1, p2, and p3, and the order of each. Note that p1 is a short-hand name; its qualified name is MapUtils.p1. We can use a qualified priority name of another DSL class in the declaration as follows:

```
1   priorities p1, p2, p3 { p1 < p2 < PredefOperators.add < p3 }
```

The order of priority names p1 < p2 indicates that an operator with the priority p1 can take as an operand an expression of an operator with the priority p2. Conversely, an operator with the priority p2 cannot take an expression of an operator with the priority p1. A priority name can be given to an operator and its operands by attaching the name enclosed by brackets. For example, the operator if-exists has the priority p1 and the right-most operand of the operator has the priority p1 in the figure. This specifies that if-exists is right-associative and if-exists can be used as an operand that

■ **Figure 6** The definition of the DSL class with operator priorities

```
1   dsl MapUtils {
2     priorities p1, p2, p3 { p1 < p2 < p3 }
3     static <K, V> void [p1] "if–exists" "(" _ "[" _ "]" ")" _ "else" _[p1]
4       (Map<K, V> map, K key, MapEntryRef<K, V> |— Void thn, Lazy |— Void els) {
5         if (map.contains(key)) thn.apply(new MapEntryRef<K, V>(map, key));
6         else els.apply(new Lazy());
7       }
8     static <K, V> Map<K, V> [p3] "{}" () { return new HashMap<K, V>(); }
9     static <K, V> V [p2] _ "[" _ "]" (Map<K, V> map, K key) { return map.get(key); }
10    static <K, V> void [p2] _ "[" _ "]" "=" _ (Map<K, V> map, K key, V value) { map.put(key, value); }
11  }
```

has a priority lower than or equal to p1. An operand without a priority name takes an expression with a higher priority than its own operator priority. Therefore, the first operand of if-exists can take an expression of the operator "it = _" but cannot take an expression of if-exists.

Operator priorities declared in a DSL class are a partial order. The total order is determined at each call site of operators. When a user imports multiple DSL classes, the compiler sorts all the priority names in the DSL classes in the declared orders. If priorities are cyclic, the compiler fails to compile a program and reports invalid operator priorities. To control operator priorities between independent DSL classes, a user can specify the order of priority names in the DSL classes at the import declaration as follows:

```
1   import dsl MapUtils { MapUtils.p2 < PredefOperators.add < MapUtils.p3 }
```

## 4  Parsing

The compiler of ProteaJ2 adopts *eager disambiguation* for fast parsing. The syntax of user-defined operators in ProteaJ2 might be ambiguous if we do not consider types and operator priorities. When syntax is ambiguous, a parser may generate a large, sometimes exponentially large, number of potentially valid ASTs for the same source code. All these ASTs except the semantically correct one are filtered out after parsing by applying semantic rules; in ProteaJ2, they are filtered out by checking types and operator priorities. If multiple ASTs are semantically correct, the language itself is ambiguous. In general, the generation of a large number of potentially valid ASTs slows down the parsing speed. Eager disambiguation is a technique to filter out semantically invalid ASTs while a parser is running. It enables faster parsing since ambiguities are resolved before the number of ASTs exponentially increases.

For eager disambiguation, ProteaJ2 uses *expected* types and operator priorities. The ProteaJ2 compiler parses a source program with a top-down parsing algorithm. Before parsing an expression, it prunes away syntax rules not in effect in the context of that expression so that the number of generated ASTs will be reduced. If the expression expects a turnstile type D ⊢ T, the compiler tries to parse it as an expression with the expected type T where the syntax of instance operators of D are in effect. Such operators are also in effect in sub-expressions of the expression. Static operators are effective everywhere if the DSL class that declares them has been imported. When the compiler tries to parse an expression with the expected type T (which is not a turnstile type), it examines only the syntax of operators that are in effect and that return the expected type T. For example, when the compiler parses an expression on the right-hand side of the assignment operator =, it examines the syntax of the operators that return a subtype of the left-hand side of the assignment. A bottom-up parsing algorithm is not suitable for such eager disambiguation. It is difficult to determine what type is expected at the expression.

Due to this approach, user-defined operators are available in an expression whose expected type is statically determined during top-down parsing. For example, user-defined operators are available in the following locations: the argument of a method call, the right-hand side of an assignment, the condition expression of an if statement, an operand of another user-defined operator, and so forth. On the other hand, they are not available at the receiver of a method call or the left-hand side of an assignment. Details of this restriction has been mentioned in our previous manuscript [12].

After parsing each (sub-)expression, if ambiguities still remain, they are resolved with heuristics; the compiler selects the AST that consumes the maximum number of characters in source code. Note that the compiler completely resolves ambiguities at every sub-expression. In other words, the compiler selects the partially optimal AST. If users define and use a dangling-else like construct, for example, the else belongs to the innermost if. This is because the compiler selects the longest match choice, if-else, for the sub-expression of outer if. This approach is similar to parsing expression grammar (PEG) [8] and packrat parsers [7]. PEG adopts the heuristics in which if there are multiple choices, the parser selects the left-most choice. The ProteaJ2 parser is based on packrat parsing supporting left recursion [28].

## 5 Evaluation

### 5.1 Time Complexity

The time complexity of our parsing algorithm is $\mathcal{O}(N * L)$ if the grammar is not left-recursive, where $N$ is the length of an input and $L$ is the number of *languages* that can possibly be used in parsing. In ProteaJ2, programmers can introduce local syntax within an expression by using a context-sensitive expression. It can be regarded as changing a language in the local scope. The compiler should use a different parser for each language, so the time complexity of our parsing algorithm is in proportion to the number of languages $L$. The base complexity of our parser is derived from the complexity of packrat parsing [7]. The time complexity of packrat parsing is $\mathcal{O}(N)$ where $N$ is the length of the input.

The number of languages $L$ can be approximated to a constant in normal cases. In some cases, however, $L$ could increase faster than linear in the length of an input $N$.

**User-Defined Operators Including Name Binding for New Language Constructs**

■ **Figure 7** Operators that could cause compilation time to increase faster than linear time

```
1   String "begin" _ "end1" (D1 ⊢ String f) {...}
2   String "begin" _ "end2" (D2 ⊢ String f) {...}
3   ...
4   String "begin" _ "endP" (DP ⊢ String f) {...}
```

Figure 7 shows a definition of operators that might increase compilation time faster than linear time. The definitions of these operators might look ambiguous. They have the same return type and the same syntax except their postfix name. Each of them has a parameter with the turnstile type like Dx ⊢ String. Such a type indicates that its argument is an expression of type String and the expression is written in a language including instance operators of Dx. If the operators in the figure are nested as follows, the number of languages exponentially increases:

```
1   begin begin begin ... begin "hello" endP endP endP ... endP
```

When the compiler reads the first begin, it cannot yet determine which of the operators in Figure 7 is used. In the worst case, the compiler should use a number of languages equal to $P$ to parse the argument of the outermost begin-end. Similarly, the compiler also cannot determine which of begin-end operators is used when it reads the next begin. Hence, for each $P$ languages for the first begin, it should examine $P$ languages to parse the expression following the second begin. The number of languages $L$ is $P^M$ in the worst case where $M$ is the number of begin operators. Since $M$ is in proportion to the length of the input, the time complexity of our parsing algorithm might be an exponent of $N$.

The cause of this problem is that the parser cannot determine which operator is used before it parses deeply nested operands. This case would rarely occur since the authors of the operators would likely design the syntax so that the operator could be uniquely determined after just reading the operator prefix such as begin.

We have conducted two micro benchmarks to examine the performance of our compiler. The first benchmark was an experiment to show that the compilation time complexity could be square or cubic time in the number of operators in unfavourable cases. The second benchmark shows that the number of operators has little impact on compilation time in normal cases. The machine used for the benchmarks had a 2.6 GHz Core i5 processor and 16 GB DDR3 memory. The operating system of the machine was Mac OS X 10.11.5.

The first benchmark uses the operators shown in Figure 7. We measured the compilation time of programs including each of the following lines while changing the number of the declared begin-endX operators $P$:

```
1   1. begin "hello, world!" endP
2   2. begin begin "hello, world!" endP endP
3   3. begin begin begin "hello, world!" endP endP endP
```

The rest of the input program includes the import declaration of the DSL class, a class declaration, and a main method. According to the discussion above, the number of the languages $L$ of code 1 is in proportion to $P$, that of code 2 is in proportion to $P^2$, and that of code 3 is in proportion to $P^3$. If this is true, the parsing time of code 2 increases faster than code 1 and the parsing time of code 3 increases faster than code 2 as the number of operators $P$ increases. Figure 8 (1) shows the result of the benchmark. The vertical axis represents the compilation time and the horizontal axis represents the number of begin-endX operators $P$. According to this figure, the parsing time of code 2 increases as the number of begin-endX operators $P$ does and it increases faster than the parsing time of code 1. The parsing time of code 3 increases faster than code 2.

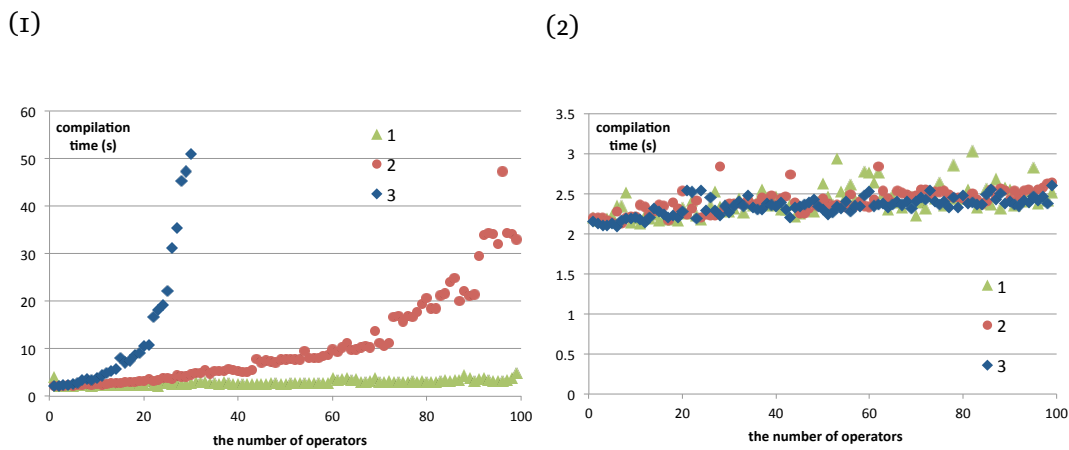

**Figure 8** The result of the micro benchmarks

The second benchmark ran different operators, whose syntax rule can be uniquely determined after just reading the prefix. Their definitions are as follows:

```
1  String "begin1" _ "end1" (D1 |— String f) {…}
2  String "begin2" _ "end2" (D2 |— String f) {…}
3  …
4  String "beginP" _ "endP" (DP |— String f) {…}
```

The input of this benchmark includes one of the following lines. Note that the prefix is not begin but beginP.

```
1  1. beginP "hello, world!" endP
2  2. beginP beginP "hello, world!" endP endP
3  3. beginP beginP beginP
4              "hello, world!" endP endP endP
```

Since the operators can be determined by their prefix, the number of the languages $L$ is a constant. Hence the parsing time of codes 1, 2, and 3 do not show notable differences even if the number of beginX-endX operators $P$ increases. Figure 8 (2) shows the result of this benchmark. The vertical axis represents the compilation time, and the horizontal axis represents the number of beginX-endX operators $P$. According to this figure, the number of beginX-endX operators $P$ has little impact on the compilation time.

■ **Figure 9** The implementation of lambda expressions

```
1   dsl Lambda <A, var: Id> {
2     static <A, B, var: Id> Function<A, B> "{" var "—>" _ "}" (Lambda<A, var> |— B body) {
3       return new Function<A, B> {
4         B apply (A a) { return body.apply(new Lambda<A, var>(a)); }
5       };
6     }
7     A var () { return value; }
8     Lambda (A a) { value = a; }
9     private A value;
10  }
```

## 5.2 Case Study

In the rest of this section, we show several use cases of our proposal. Although ProteaJ2 does not currently support anonymous classes, we use anonymous classes for simplifying a program shown in this section.

**Lambda Expressions**

We can implement lambda expressions in ProteaJ2. Lambda expressions are a common feature of modern programming languages and they have also been introduced in Java. They create an anonymous function as a first-class object. Figure 9 shows the implementation of lambda expressions. The operator "{" var "->" _ "}" takes a parameter name as a generic name var. The operator var (declared at line 7) is available as the operand of the operator "{" var "->" _ "}" since the operand type is the turnstile type Lambda<A, var> ⊢ B. Here, the operator "{" var "->" _ "}" and the operator var refer to the same generic name. Therefore, this definition correctly expresses name binding of a lambda expression. The value bound to the name is given when the apply method of the returned object (defined in line 4) is called. When we import this definition, we can write the following code snippet:

```
1   Function<String, String> f = { s —> s + "!" };
```

We can also implement a variant of lambda expressions in ProteaJ2; although a typical lambda expression consists of a single function, we can implement a lambda-like expression that consists of multiple functions. Figure 10 shows a DSL class that implements such syntax for KeyListener. It can be used as follows:

```
1   Component c = …;
2   c listens key {
3     pressed —> System.out.println(key name + " pressed!")
4     released —> System.out.println(key name + " released!")
5     typed —> key_typed(key)
6   };
```

This code registers a KeyListener that prints a key name when the key is pressed or released and also invokes the key_typed method when a key is typed. On the right-hand side of each ->, key acts as a variable of the type KeyEvent. Note that key

**Figure 10** The definition of the DSL class that implements syntax like lambda expressions for KeyListener

```
1   dsl KeyListenerDSL {
2     static void
3     _ "listens" "key" "{"
4       "pressed" "—>" _
5       "released" "—>" _
6       "typed" "—>" _
7     "}"
8     (Component c, KeyListenerDSL |— Void f1,
9      KeyListenerDSL |— Void f2, KeyListenerDSL |— Void f3) {
10        c.addKeyListener(new KeyListener {
11          void keyPressed (KeyEvent e) { f1.apply(new Key(e)); }
12          void keyReleased (KeyEvent e) {...}
13          void keyTyped (KeyEvent e) {...}
14        });
15    }
16    KeyListenerDSL (KeyEvent e) { keyEvent = e; }
17    KeyEvent "key" () { return keyEvent; }
18    String "key" "name" () { return KeyEvent.getKeyText(keyEvent.getKeyCode()); }
19    private KeyEvent keyEvent;
20  }
```

name acts as a variable of the type String. We can provide multiple accessors to obtain the same underlying value as values of different types.

**The Loan Pattern**

The loan pattern is a design pattern to avoid explicitly releasing resources. It is useful for expressing I/O processing. Several languages support the loan pattern as a language construct, for example, using in C# and try-with in Java. The following code uses try-with in Java:

```
1  try (BufferedReader reader = new BufferedReader(new FileReader(file))) {
2    System.out.println(reader.readLine());
3  }
```

It is guaranteed that reader is closed at the end of the try statement.

In ProteaJ2, we can implement this as if it is a built-in language construct. Figure 11 shows the definition of the DSL class TryWith, which implements the language construct try-with. Similar to a lambda expression, the operator try declares the name id and this name is available in the body of try. The name id is bound to the given resource from the first operand of try. It is guaranteed to release the resource at the end of the try since the close method is called in the finally clause from line 7 to line 9.

**Hiding Resources**

In ProteaJ2, we can develop a much safer DSL than try-with for managing resources. With the loan pattern, users directly access a resource object, which is dangerous since users can assign the resource object to a variable in an external scope. In ProteaJ2,

**User-Defined Operators Including Name Binding for New Language Constructs**

■ **Figure 11** The definition of the DSL class for implementing try-with

```
1   dsl TryWith <R extends AutoCloseable> {
2     static <R extends AutoCloseable, id: Id> void "try" "(" R id "=" _ ")" "{" _ "}"
3       (Lazy |─ R resource, TryWith<R> |─ Void body) throws Exception {
4       R r = resource.apply(new Lazy());
5       try { body.apply(new TryWith<R>(r)); }
6       finally { if (r != null) r.close(); }
7     }
8     R id () { return resource; }
9     TryWith (R r) { resource = r; }
10    private R resource;
11  }
```

we can encapsulate the resource object as a field of a DSL class and provide instance operators only within a context-sensitive expression. Figure 12 shows the definition of such the DSL class that encapsulates a `BufferedReader` object.[3] We can write the following code snippet if we import that DSL class:

```
1   open "file.txt" {
2     while (has next) { System.out.println(read line); }
3   };
```

The users cannot directly access a `BufferedReader` object but they can use operators has next or read line in the body of the operator open. Here, while statement appears in the body of the operator open. In ProteaJ2, a turnstile type like D ⊢ Void is a special type that can take a block of statements as its argument. Since the body of the operator open has the turnstile type FileRead ⊢ Void, it can take a block that contains while statement.

**Dynamic Scoping**

ProteaJ2 provides a requires clause for implementing a method or an operator that is only available within an operand of a user-defined operator. A method or an operator with the requires clause becomes available if all the specified types in the required clause are assumptions (i.e., within a context-sensitive expression of a turnstile type containing the specified DSL type as the assumption). The following code uses a method with a requires clause:

```
1   List<String> allLines = open "file.txt" { getLines() };
2   List<String> getLines() requires FileRead {
3     List<String> lines = new ArrayList<String>();
4     while (true) {
5       String line = read line;
6       if (line == null) return lines; else lines.add(line);
7   }}
```

---

[3] Technically, this code does not work in the current version of ProteaJ2 since we implement a turnstile type by using java.util.function.Function. ProteaJ2 does not allow checked exceptions, so the value of the turnstile type FileRead ⊢ Void cannot throw checked exceptions.

■ **Figure 12** The definition of the DSL class that encapsulates BufferedReader

```
1  dsl FileRead {
2    static void "open" _ _ (String fileName, FileRead |— Void body) throws IOException {
3      BufferedReader reader = new BufferedReader(new FileReader(fileName));
4      try { body.apply(new FileRead(reader)); }
5      finally { if (reader != null) reader.close(); }
6    }
7    String "read" "line" () throws IOException {
8      String res = nextLine;
9      this.nextLine = reader.readLine();
10     return res;
11   }
12   boolean "has next" () { return nextLine != null; }
13   FileRead (BufferedReader reader) throws IOException {
14     this.reader = reader;
15     this.nextLine = reader.readLine();
16   }
17   private BufferedReader reader;
18   private String nextLine;
19 }
```

The getLines method is available only within the second operand of open. In the method body of getLines, the operator read line is available since FileRead is the assumption there. We can regard a required clause as an emulation of a dynamic scope since getLines can access the operators available in the caller's scope.

**Pattern Matching**

In ProteaJ2, programmers can implement expressions involving more complex name binding. For example, they can implement a simple pattern matching mechanism, which is a common language construct in functional programming languages such as Haskell and Scala. Figure 13 shows part of its definition. The complete definition is available from our github repository.[4] It can be used as follows:

```
1  String s = "hello, world" match {
2    | "hello" + xs => "goodbye" + xs
3    | Empty => "empty string"
4    | otherwise => otherwise
5  };
```

This code displays "goodbye, world" to the user. In this code, xs and otherwise act as variables. The names of the variables are bound by the operator defined at line 16 in Figure 13. Note that a variable name is restricted here to be an identifier that starts with a lower case letter. This restriction is borrowed from the pattern matching mechanism in Scala.

---

[4] https://github.com/csg-tokyo/proteaj2

**User-Defined Operators Including Name Binding for New Language Constructs**

■ **Figure 13** Part of the definition of pattern matching

```
1   dsl MatchDSL {
2     static <T, R> R _ "match" "{" _+ "}" ( T t, Case<T, R>... cases ) {
3       for (Case<T, R> c : cases) {
4         Optional<R> result = c.apply(t);
5         if (result.isPresent()) return result.get();
6       }
7       return null;
8     }
9     static <T, R> Case<T, R> "|" _ "=>" _ ( Matcher0<T> m, Lazy |— R f ) {
10      return new Case<T, R> ({ t —> m.ifMatch(t, f) });
11    }
12    static <T, A, name: Id_Lower, R> Case<T, R>
13     "|" _ "=>" _ ( Matcher1<T, A, name> m, Var<A, name> |— R f ) {
14      return new Case<T, R> ({ t —> m.ifMatch(t, f) });
15    }
16    static <T> Matcher0<T> "_" () { return new Matcher0 <T> ({ t => Boolean.TRUE }); }
17    static <A, name: Id_Lower> Matcher1<A, A, name> name () {
18      return new Matcher1 <A, A, name> ({ a —> Optional.<A>of(a) });
19    }
20    static Id_Lower _ _* (LowerCase lc, Letter... letters) {
21      return new Id_Lower(lc, letters);
22    }
23  }
24  dsl StringMatchers {
25    static Matcher0<String> "Empty" () {
26      return new Matcher0 <String> ({ t —> Boolean.valueOf(t.isEmpty()) });
27    }
28    static <A, name: Id_Lower> Matcher1<String, A, name>
29     _ "+" _ ( String prefix, Matcher1<String, A, name> m ) {
30      return new StringPrefixMatcher <A, name> (prefix, m);
31    }
32  }
33  dsl Var <A, name: Id_Lower> {
34    Var (A a) { this.a = a; }
35    A name () { return a; }
36    private A a;
37  }
```

## 6  Related Work

SugarJ [6] is a programming language that supports user-defined syntax extensions based on macros. Programmers can define any context-free syntax in SDF [10] and any AST transformation rules in Stratego [27]. The defined syntax and AST transformation rules are modularized as a *sugar library*. Users can import the sugar library and use the defined syntax. The users can also use multiple sugar libraries in the same source file. Unfortunately, the compiler does not have the capability to handle ambiguous syntax. This limitation breaks the composability of libraries, as several libraries cannot be used together. Several languages such as Nemerle [23] also support user-defined syntax extensions based on macros but their capabilities are more limited. In the case of Nemerle, the users can only define the syntax where the first token is an identifier. Moreover, terminal symbols must be selected from a specific set of identifiers and operators.

Type-specific languages (TSLs) [17] and typed syntax macros (TSMs) [18] are a language feature that supports composable user-defined syntax extensions. Here, composable means that users can use several syntax extensions in the same source file. When the users use the multiple syntax extensions together, syntactic ambiguities might occur. The compiler resolves such syntactic ambiguities by the expected type information. TSLs are similar to our previous work, ProteaJ, because they can give custom syntax to a function. TSMs are similar to SugarJ since they are based on syntactic macros, but they are more limited since each macro is statically typed. TSLs and TSMs can implement various syntax extensions and the extensions are safe to compose, however, they do not have the capability to define custom name binding. Since TSLs are just a function with custom syntax, they cannot extend name binding rules. TSMs are focused on the composability of syntactic macros. Hence they only handle hygienic macros [14], which are syntactic macros that do not affect name binding. Metamorphic syntax macros [3] are also syntactic macros that guarantee type safety. They do not affect name binding for the same reason as TSMs. TSLs and TSMs partially adopt eager disambiguation to parse a program in acceptable time. They provide several literal forms for using user-defined syntax. The compiler initially skips parsing such literal forms until after type checking the surrounding code. When parsing the literal, only syntax extensions that return the expected type in that position are considered.

Mixfix operators [5] are another kind of user-defined operator. They were adopted in several languages such as Coq [25] and Isabelle [19]. Unlike ProteaJ2, to reduce syntactic ambiguities, mixfix operators only allow users to use particular forms of syntax: prefix, postfix, infix, or outfix syntax. Several languages also allow programs to define operators without a name part. Since mixfix operators are user-defined operators, it is difficult to use them to implement language constructs involving name binding. Generalized method names [11] extend multi-part method names, which are found in several languages such as Smalltalk, so that they take regular expressions instead of a fixed name. They can express various syntax similar to user-defined operators. They also have the same problem with name binding as user-defined operators since they just extend the syntax of method calls.

Implicit parameters in Scala [16] are a similar idea to our proposal. An implicit parameter is a method parameter whose argument is given implicitly if the argument is not explicitly specified at the call site of the method. The actual argument is found among implicit values visible in the current scope. The difference is that the arguments of our context-sensitive expression are given from the outer method or operator. Implicit parameters enable a context-sensitive expression to express syntax available only within a local scope. We propose a turnstile type for expressing such an expression. The Glasgow Haskell Compiler (GHC) has an extension named ImplicitParams based on [15]. It enables programmers to use variables with a dynamic scope in a type safe fashion. ImplicitParams is similar to our requires clause mentioned in section 5.2.

Kotlin provides "function literals with receiver", which enables calling a function literal with a specified receiver object. It is used for writing type-safe Groovy-style builders in Kotlin. Function literals with receiver provides similar capability as a context-sensitive expression. A major difference from our proposal is Kotlin does not aim to implement name binding, so it does not have the functionality to support syntax extension with name binding such as DSL classes and generic names. instance_eval in Ruby is also related to context-sensitive expressions. instance_eval evaluates the given block as if it is an instance method of the given receiver. instance_eval can be regarded as a context-sensitive expression implemented in a dynamically typed language. Since name binding is resolved at runtime, instance_eval is not safe.

Type-oriented island parsing [22] is a parsing technique to efficiently parse a source program. It is an extension of island parsing [24], which is an efficient bottom-up parsing algorithm using heuristics. The type-oriented island parsing performs eager disambiguation for a completed child AST by using type information. The paper also shows that they have implemented a syntax extension system with name binding (called variable binders in that paper). To express name binding, they adopted labeled symbols [13] and a scoping construct [4]. Labeled symbols are similar to generic names in our system, and a scoping construct is similar to the left-hand type of a turnstile type. However, these constructs express a variable naively and the variable is not encapsulated. They cannot change the syntax of names or publish a different interface for accessing a variable. In other words, a DSL class is a module system that reorganizes and generalizes labeled symbols and a scoping construct to integrate them with an object-oriented language. Furthermore, as far as we know, it is not clear whether or not the type-oriented island parsing can be applied to scannerless parsers since type-oriented island parsing uses heuristics for parsing tokens. If it cannot be applied to scannerless parsers, programmers cannot extend literal syntax.

## 7 Conclusion

This paper proposes a variant of a lambda expression, named a *context-sensitive expression*, for implementing user-defined language constructs with custom name binding. A context-sensitive expression looks like a normal expression, but it implicitly takes parameters. These parameters are accessed through public members such as methods and operators of a parameter object instead of through parameter names. We have developed ProteaJ2, which is a programming language based on Java but supports context-sensitive expressions. ProteaJ2 supports *turnstile types*, *DSL classes*, and *generic names* to express context-sensitive expressions for syntax extension with name binding. The parser of ProteaJ2 adopts *eager disambiguation* using expected types to efficiently parse a program. We showed several micro benchmarks to illustrate that compilation time is acceptable. We also showed several case studies to demonstrate our system can implement various language constructs with custom name binding.